\documentclass[acmtog]{acmart}
\usepackage{todonotes}
\usepackage{color,soul}

\usepackage{pythonhighlight}
\usepackage{amsmath}
\usepackage{float}
\usepackage[bottom]{footmisc}

\AtBeginDocument{%
  \providecommand\BibTeX{{%
    \normalfont B\kern-0.5em{\scshape i\kern-0.25em b}\kern-0.8em\TeX}}}

\setcopyright{acmcopyright}
\copyrightyear{2018}
\acmYear{2018}
\acmDOI{10.1145/1122445.1122456}

\acmJournal{TOG}
\acmVolume{37}
\acmNumber{4}
\acmArticle{111}
\acmMonth{8}

\begin{document}

\title{BOPI: A Programming Interface For Reuse Of Research Data Available On DSpace Repositories}

\author{Andrew Edric Tristan}
\authornote{Both authors contributed equally to this research.}
\email{andrew.tristan@campus.tu-berlin.de}
\orcid{1234-5678-9012}
\author{Vinicius Woloszyn}
\authornotemark[1]
\email{woloszyn@tu-berlin.de}
\affiliation{%
  \institution{Technische Universität Berlin}
  \streetaddress{P.O. Box 1212}
  \city{Berlin}
  \state{Berlin}
  \country{Germany}
  \postcode{43017-6221}
}

\author{Ben Kaden}
\affiliation{%
  \institution{Institut für Bibliotheks- und Informationswissenschaft der Humboldt-Universität zu Berlin}
  \city{Berlin}
  \country{Germany}
}
\renewcommand{\shortauthors}{Tristan and Woloszyn, et al.}


\begin{abstract}
A recent study showed that more than 70\% of researchers fail to reproduce their peers's experiments and more than half fail to reproduce their own experiments. Obviously, from a perspective of scientific quality this is a more than unsatisfying numbers. One approach to mitigate this flaw lies in the transparent provision of relevant research data to increase the base of available material to evaluate and possibly reconduct experiments. However, such data needs to be presented and accessed in a findable and purposefully usable way. In this work, we report the development of a programming interface to enhance findability and accessibility of research data (available in DSpace systems) and hence reproducibility of scientific experiments with data. This interface allows researchers to (i) find research data in multiples languages trough automatic translation of metadata; (ii) display a preview of data without download it beforehand; (iii) provide a detailed statistics of the data with interactive graphs for quality assessment; (iv) automatic download of data directly from Python-based experiments. Usability tests revealed that this interface improves the effectiveness, efficiency and satisfaction during the reuse of research data.
\end{abstract}

\ccsdesc[500]{Computer systems organization~Embedded systems}
\ccsdesc[300]{Computer systems organization~Redundancy}
\ccsdesc{Computer systems organization~Robotics}
\ccsdesc[100]{Networks~Network reliability}

\keywords{datasets, neural networks, gaze detection, text tagging}

\maketitle

\section{Introduction}
Research data is any form of distinctly codefied material that has been collected, observed, generated or created to produce original research results. Some common examples of Research Data are textual documents, spreadsheets, computer notebooks, models, algorithms, scripts, etc. The nature and form of this data depend on disciplinary and methodological factors, which can vary significantly across respective communities of research. 

A challenge for infrastructure services and especially data repositories lies in the form of an appropriate presentation and provision of access of very heterogenous materials and scenarios of access and usage. Hence, solutions for this challenge need to be rather generic, but still ensure appropriateness for scholarly and scientific use. Basic parameters of scientific work and communications, such as distinct identifiability, as well as citeability, need to be guaranteed. 

In terms of regulating the provision of research data the so called FAIR-principles aim to guide researchers towards a more open and extensive state of practice \cite{wilkinson2016fair}. The central issue however in many disciplines remains the general lack of a systematic availablity of relevant Research Data for both reasons: evaluation and quality assessment of research results, as well as reuse. An additional aspect is a possible gain of merits through data publications which at least in some settings gets increasingly relevant in terms of impact and performance measurement for individual researchers and teams and institutions.  

Over recent years, several incentives to increase data sharing through Data Repositories have been established. Many funding agencies, journal publishers, academic institutions and research organizations across the globe have implemented new strategies for research data sharing, often to comply with new governmental directives \cite{attitudes}. However, while recent developments have produced substantial growth in data availability, the data repositories tend to focus on functional aspects and do not regard usability aspects, such as epistemological factors, i.e., the logic of scholarly work. For example, recent studies have concluded that a prerequisite for reuse of research data is also the utilization of appropriate data gathering methods, organization tools, and workflows early on in the research life-cycle~\cite{attitudes,improve}.

In this study, we present BOPI\footnote{Berlin Open Science Python Interface}, an open-source tool to support researchers early in their experiments with data. For example, while looking for suitable research data to perform (or reproduce) an experiment, scientists can be exposed to a few error-prone tasks, such as search for the data, translation of metadata, visualization of the data, manual download, etc. The goal of BOPI is enable researches to automatize main error-prone tasks and improve the reuse of research data, as well as Effectiveness, Efficiency and Satisfaction. The main features of BOPI are:

\begin{enumerate}
    \item A Program Interface for searching research data in multiple DSpace repositories, at the same time;
    \item Automatic translation of metadata;
    \item Quick preview of tabular data, including head and bottom of the dataset, without the need of downloading it;
    \item automatic integration and download of data directly from Python programs or Jupyter Notebooks\footnote{Jupyter Notebooks are open source web application that allow researchers to combine software code, visualize data or explore the data (https://jupyter.org/)};
    \item Automatic Data transformation to several formats, e.g., CSV, JSON, XLS, and etc.
\end{enumerate}

We therefore hypothesize that:

\begin{itemize}
    \item \textbf{H1} the use of a Programming Interface that enhances DSpace Systems' features improves the Efficiency of researchers in reproducing experiments;
    \item \textbf{H2} the use of a Programming Interface that enhances DSpace Systems' features improves the Effectiveness of researchers in reproducing experiments;
    \item \textbf{H3} the use of a Programming Interface that enhances DSpace Systems' features improves the Satisfaction of researchers in reproducing experiments; 
\end{itemize}

In order to test our hypotheses, we have developed a prototype of BOPI. The decision to choose Python for developing this tool is motivated mainly because (i) our target audience are technicians who have some basic programming skills; and (ii) because Python is considered as main Open-Source language for performing experiments \cite{python}; and (iii) DSpace Systems\footnote{https://duraspace.org/dspace/} are employed widely in different research institutions over the globe, as well as in Berlin University Alliance\footnote{https://www.berlin-university-alliance.de/}, where this research is conducted. Additionally, the programmatic way of accessing Research Data in multiple Repositories is an important step for the recent efforts to an automatic processing and interpretation of Scientific Work, such as populating of an Open Research Knowledge Graph \cite{jaradeh2019open}.

We describe Background and related work in Section \ref{Background and related work}. Our proposed method is introduced in Section \ref{proposed}, followed by a experiment design in section \ref{ExpDesign}. Afterward, the results of the evaluation will be presented in section \ref{results}. Moreover, some use cases are described in section \ref{useCase}. Lastly, we close this study with the discussion in section \ref{discussion} followed by the conclusion and future works in section \ref{conclusion}.

\section{Background and Related Work}
\label{Background and related work}
\subsection{Preliminaries}
To clarify the ideas and objectives that this document addresses, we first introduce some essential concepts in this section used in the remainder of this work.

\begin{itemize}
    \item Data: the term refers to a set of values of qualitative or quantitative variables about one or more persons or objects ~\cite{data}.
    \item Research Data: is any information that has been collected, observed, generated or created to validate original research findings ~\cite{reData}.
    \item Data set: a data set is a collection of data. In the case of tabular data, every column of a table represents a particular variable, and each row corresponds to a given item of the data set ~\cite{dataset}.
    \item Data Repository: it is a place that holds data, makes data available to use, and organizes data in a logical manner ~\cite{repo}.
    \item Metadata: Structured information that describes or inform about the data, such as the format, title, abstract or authors name ~\cite{metadata}.
    \item Computer Notebooks: The open-source web application that allow the users to manage, create, share documents. The documents may contain the code, text, visualization that can be used for machine learning, data cleaning and much more~\cite{JP}.
\end{itemize}

\subsection{Usability}
\label{evaluation}

The term Usability commonly refers to the quality factors of a product or system. It is applied to explicate those properties of a product that influence how well a product responds to user interaction and how well users can achieve their usage goals. A high degree of usability means that users can achieve their particular usage goals with manageable effort, few or no disturbances, and within a clear and understandable frame for navigation and interaction; i.e., the design and the functionality of a product or system are as closely aligned as possible with the expectations, goals, as well as usage competencies of its intended user groups ~\cite{usability_definition}.  

How usability is measured depends on the usability models and metrics chosen. In this work, we applied ISO 9241-11 Ergonomics of human-system interaction — Part 11: Usability: Definitions and concepts where we focus on how users effectively and efficiently access, preview, and reuse the data from the DSpace repository as well as their satisfaction as they aim to achieve their usage goals. The following section will further explain the properties of Effectiveness, Efficiency, and Satisfaction regarding the chosen metrics.

The measurement for the usability of systems or products can be achieved through 3 standard observable and quantifiable vital metrics that have been recommended by ISO 9241-11\footnote{https://en.wikipedia.org/wiki/ISO\_9241} with effectiveness(Task completion rates), efficiency (Completion time), and satisfaction (The experience responded by user). This standard model identifies usability aspects and the context of usage elements such as users, tasks, equipment, and environments~\cite{usability1998} to be applied during the specification, design, and usability evaluation and developed by a group of experts in the field of Human-Computer-Interaction~\cite{Usabilitymodels}. 

\subsubsection{Metrics for assessment of effectiveness}
The usability literature defines effectiveness in terms of accuracy and completeness~\cite{usability2018}. For example, if the goal is to accurately preview and download datasets in a specified format, accuracy could be measured by the percentage of the user correctly and completely reach the goal, i.e. completion rate.
Effectiveness can be measured using either a completion metric as well (percentage of tasks accomplished) ~\cite{metrics} and / or a error metric (average error rates during the process of use). While a flawless usage experience is hardly achievable, a decrease of potential error sources is desired ~\cite{metrics}.

\subsubsection{Metrics for assessment of efficiency}
Efficiency is related to the utilization of resources can include time and effort. For example the time to perform the task ~\cite{usability2018}.
For the metrics of efficiency, there are some commons metrics such as:
\begin{itemize}
    \item Time-Based efficiency: The time where participants successfully finish the tasks ~\cite{metrics}.
    \item Documentation or help's use frequency ~\cite{Usabilitymodels}.
    \item Time To spent on error ~\cite{Usabilitymodels}.
\end{itemize}

\subsubsection{Metrics for assessment of satisfaction}
The satisfaction measures the attitude of the users while using the product. Using a provided questionnaire such as:
\begin{itemize}
    \item SUS (System Usability Scale): it consists of 10 questions. It is a cheaper and quicker way to evaluate practically any kind of system.
    \item  SUPR-Q (Standardized Use experience Percentile Rank Questionnaire): it consists of 8 Questions and focuses more on evaluating the websites.
    \item  QUIS (Questionnaire For User Interaction Satisfaction): Like SUS but more complex, this is a tool to assess user satisfaction ~\cite{quis}.
    \item  SUMI (Software Usability Measurement Inventory): SUMI is recommended if the main point is the user's satisfaction and there is enough budget allocated ~\cite{metrics}.
\end{itemize}

Among those mentioned questionnaires, SUS is favored to have a more accurate result, there is not a large allocated budget, and consists of an ideal scale for the participants for the usage in simple evaluation ~\cite{metrics}.

\subsection{Related Work}

Over recent years, there has been an increasing interest in improving the reuse of research data. For example, Curty et al. ~\cite{attitudes} have investigated the attitudes and norms that affect the reuse of research data on DataOne\footnote{dataone.org} repositories. They have analyzed the relationship between the beliefs and attitudes of scientists towards data use. The study showed that the perceived efficacy and efficiency of data repositories are strong predictors for reusing data. Additionally, they have classified the indicators into four distinct groups:

\begin{itemize}
    \item A great deal of trust in data producers and heavy reliance on the methods and techniques they employed to obtain, organize, and code the data.
    \item Comprehensive metadata (to support the correct interpretation of the data).
    \item Some demonstrations to help reduce the initial barrier of data reuse by demonstrating more practically and palpably its value, such as YouTube video case studies of data reuse or Jupyter notebooks demonstrating the process of reusing data.
    \item Formal citation of datasets is key to leveraging data sharing and legitimizing the research data's reuse at a time.
\end{itemize}

A taxonomy of usability issues of research data repositories was proposed by Volentine et al. ~\cite{improve}. They have performed usability tests on Data Observation Network for Earth (DataONE) repositories, which revealed several issues that were classified into four distinct groups:
\begin{itemize}
    \item Semantic issues dealing with unfamiliar words and phrases that confused the user.
    \item Technical usability issues dealing with functionality and navigability of tools itself.
    \item Structural usability issues encompass the problems with information architecture, for instance: the ability to download data and metadata as one compressed file.

    \item Aesthetical usability issues dealing with the display of the content itself.
\end{itemize}

\section{Proposed Solution: BOPI}
\label{proposed}

Previously studies have pointed out the need for improvement of reproductivity of scientific experiments\cite{attitudes}. Therefore, this work proposes a middleware that automatizes the data-gathering task since it is an initial and essential phase for experimenting. Moreover, there has been an increasing interest in processing and interpretation of scientific content\cite{improve}, and the development of an intuitive way to access the data would enable large-scale processing of this information.

There are many different repositories and possible ways to perform experiments. This work addresses a particular research community that uses data from Dspace repositories and performs experiments using Python-based language such as Computer Notebooks. This decision is mainly taken due to Dspace Systems are extensively adopted by research institutions and the institution where this research is caring on. The use of Dspace Systems also enables us to carry in-depth experiments using a real data repository. Additionally, due to modular architecture, it is possible to reuse this middleware in future projects involving research data processing. 

Besides automatizing data-gathering, it allows users to search for data using features that are not possible when using standard Dspace System web interfaces, such as:

\subsection{Looking for data on Multiple Data Repositories}
Using BOPI, it is possible to search for research data in multiple repositories at the same time. It is by default configured to look at official research data repositories of the Berlin University alliance. In this work, we only use the Depositonce repository as an example. Let us say that the user is looking for data regarding Temperature and Humidity.

\begin{lstlisting}[style=mypython,label={code:search},caption={Code sniped   of how to Searching data for Temperature and Humidity data}]
import bopi
bopi.search("Temperature and Humidity")
\end{lstlisting}

\begin{figure*}
\includegraphics[width=\linewidth]{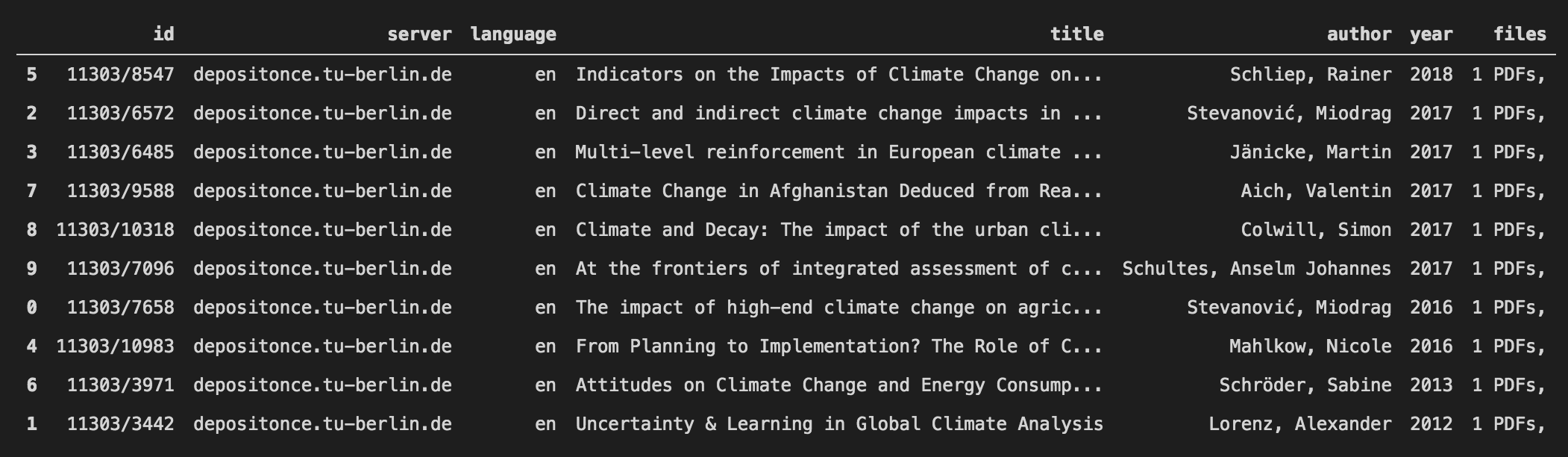}
\caption{Return of the command search using the string "Temperature and Humidity Measurements in Gardens" as parameter: Research Data available at Depositonce TU Berlin}
\label{fig:results_search2}
\end{figure*}

The results of the command \ref{code:search} is a table depicted on Figure \ref{fig:results_search2}, where the columns are:

\begin{itemize}
    \item \textbf{ID}: A unique identifier generated by DSpace system.
    \item \textbf{Server}: This provides information about the location of the repository.
    \item \textbf{Language}: It shows the languages available for the document.
    \item \textbf{Title}: The name of each documents.
    \item \textbf{Author}: It describes the author's name of the item or document.
    \item \textbf{Year}: The year of publication.
    \item \textbf{Files}: It describes the number and format inside the item or document.
\end{itemize}

\subsection{Automatic Translation Of Metadata}
Sometimes, the authors provide metadata in only one language, making it difficult to be found. It can be an obstacle to the internationalization of research. We have employed Google Translation API for making an automatic translation of the available metadata. 
Introduce the problem of short and domain-specific translation. 

\begin{lstlisting}[style=mypython,caption={Automatic Translation of Metadata to English}]
import bopi
bopi.search("Temperature and Humidity in Gardens", 
translate_to="en")
\end{lstlisting}

\subsection{Preview of different types of Tabular Data}

Before the users begin to preview, describe or download the data, they should get the list the available data from one repository by using followong syntax:
\begin{lstlisting}[style=mypython,caption={Get the list of  the Data}]
import bopi
bopi.repository("11303/10989.2").datasets()
\end{lstlisting}

Checking out data's content can be a time-consuming task. This can happens due to a slow download rate of a large file or the time expended unzipping and looking for the right application to open it. Therefore, \textbf{BOPI} provides a quick preview of the data without neither download nor the use of a third-party application.

\begin{lstlisting}[style=mypython,caption={Preview of  the Data}]
import bopi
bopi.repository("11303/10989.2").datasets()
.get('name_of_file'.csv).preview()
\end{lstlisting}

\subsection{Data Visualization}
The other feature is Data profiling, it is the process to explore and analyze data from existing resource. 

\begin{lstlisting}[style=mypython,caption={Exploring the Data},label={command:data_describe}
]
import bopi
bopi.repository("11303/10989.2").datasets()
.get('name_of_file.csv').describe()
\end{lstlisting}

\begin{figure}[H]
  \includegraphics[width=\linewidth]{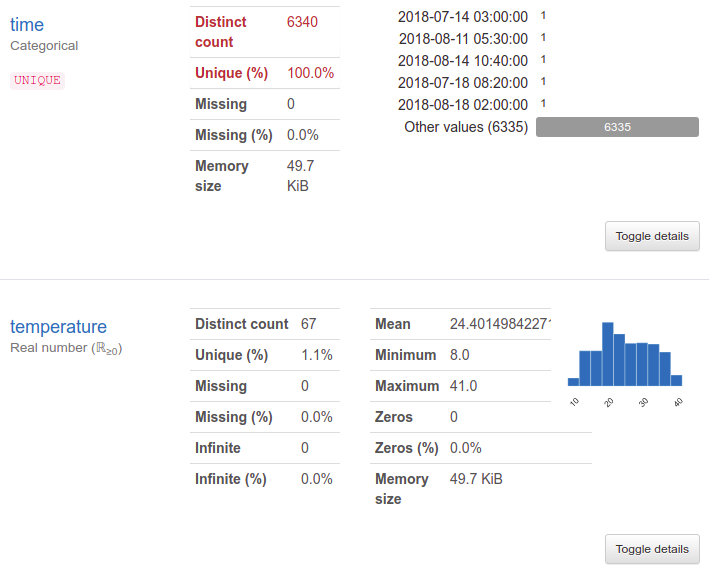}
  \caption{Example of data profiling that contains information about variable Time and Temperature.}
  \label{fig:data_describe}
\end{figure}

As one of the outputs of command \ref{command:data_describe}, Figure \ref{fig:data_describe} describes the distribution of two variables available on the data set, namely "time" and "temperature". It gives an overview of a distinct values,  missing values and the distribution of the values.

\subsection{Downloading of Research Data}
One benefit of automatic download of data is integration with experiments, improving its reproduction. 

\begin{lstlisting}[style=mypython,caption={Automatic Download certain Data from the repository}]
import bopi
bopi.repository("11303/10989.2").datasets()
.get('name_of_file.csv').download("/data")
\end{lstlisting}

\begin{lstlisting}[style=mypython,caption={Automatic Download of all Data from one repository}]
import bopi
bopi.repository("11303/10989.2").download("/data")
\end{lstlisting}

\section{Experiment Design}
\label{ExpDesign}

\subsection{Pre-evaluation}

One of the first steps to do evaluation planning for the test. The purpose of the plan itself is to get well prepared before conduct the real evaluation that will help us improve our tool and to prepare the requirements that include:
\begin{itemize}
    \item The scenarios we will use.
    \item The scope and number of participants.
    \item How we conduct the test.
    \item What kind of questionnaire will be asked to the participants.
    \item What to evaluate.
    \item What kind of documentation to explain the tools
\end{itemize}

We have invited two people who have programming and data science experiences to execute the tasks and questionnaire for the participants. Moreover, the documentation has been prepared and ready to use in our GitHub link: \href{https://github.com/TUB-NLP-OpenData/usability_research_data/}{usability research data}.

\subsection{Participants}
Our target group will encompass the students with background of programming, such as: 
Computer Science (BSc and MSc), Computational Neuroscience, Wirtschaftsinformatik, Computational Engineering Science.

\begin{table}[h!]
\begin{tabular}{|l|l|}
\hline
\textbf{Role} & \textbf{Total}   \\ \hline
Computer Science & 9  \\ \hline
Computer Engineering & 2  \\ \hline
Developer in automotive & 1   \\ \hline
Business Intelligence Manager & 1   \\ \hline
IT Consultant & 1   \\ \hline
Software Developer & 1   \\ \hline
Wirtschaftsinformatik & 1   \\ \hline
Software tester & 2   \\ \hline
Geodesy and Geo information Science & 2   \\ \hline
\end{tabular}
\caption{Description of the participants}
\label{tbl:particip}
\end{table}

\subsection{Evaluation Metrics}

We want to measure the Effectiveness, Efficiency, and Satisfaction of reuse research data available on DSpace repositories with and without the help of our tool. Therefore, the evaluation of our tool will be performed such as ~\cite{metrics}:

 \begin{itemize}
        \setlength\itemsep{1em}
        \item Effectiveness is typically measured by calculating the completion rate.
        In this case, the users will perform the tasks, such as searching the data, profiling the data, or downloading the data. In the end, we will collect the number of tasks completed successfully to achieve the percentage of efficiency. The way to calculate it can be seen in figure \ref{fig:completionRate}.
        
        \begin{figure}[h]
            \centering
            \[ Effect = \frac{Number \, of\,  tasks\,  completed\,  sucecessfully}{Total \, number\,  of tasks\,  undertaken} \times 100\% \]
            \caption{Calculation of completion rate}
            \label{fig:completionRate}
        \end{figure}

        \item Efficiency will involve the measurement based on participants' time to complete a task and its overall efficiency.
        Here, we will calculate the average time need to finish all tasks in seconds.

        \begin{figure}[h]
            \centering
            \[ TBE = \sum_{j=1}^{R}\sum_{i=1}^{N} \frac{\frac{n_{ij}}{t_{ij}}}{NR} \]
            \caption{Timed-Base Efficiency}
            \label{fig:TBE}
        \end{figure}
        
        \begin{figure}[h]
            \centering
            \[ ORE = \frac{\displaystyle\sum_{j=1}^{R}\sum_{i=1}^{N} n{_{ij}} t_{ij}}{\displaystyle\sum_{j=1}^{R}\sum_{i=1}^{N}t{_{ij}}} \]
            \caption{Overall Relative Efficiency}
            \label{fig:ORE}
        \end{figure}
        
Where N in the total number of tasks and R = The number of users.$  n_{ij} $ is the result of task i by user j; if the user successfully completes the task, then Nij = 1, if not, then Nij = 0. $  t_{ij} $ is the time spent by user j to complete task i. If the task is not successfully completed, then time is measured till the moment the user quits the task.

        \item Satisfaction can be measured by giving the formalized questionnaire to participants after they finish the tasks to measure their satisfaction level. The template questions will be given to the users, and they will have ranked each of the questions from 1 to 5 based on their level of agreement, with odd-numbered defines positive questions and even-numbered defines negative questions. The content of the questionnaire will be the questions that measure the satisfaction of the users, e.g.: 
        
        \begin{enumerate}
            \setlength\itemsep{0.2em}
            \item I think that this tool reproduce the data effectively.
            \item I found the tool unnecessarily complex.
            \item Remembering names and the use of commands is easy.
            \item I think that I would need the support of a technical person to be able to use this tool. 
            \item I felt very satisfied using the tool.
            \item I found the tool very cumbersome to use.
            \item Using the tool would improve my productivity.
            \item I thought there was too much inconsistency in this tool.
            \item Using the tool in my job would enable me to accomplish tasks more quickly and efficiently.
            \item I found the various functions in this tool were "not" well integrated.
        \end{enumerate}

\end{itemize}

\subsection{Scenarios}
For the scenario, we have divided it into two parts; we have decided to use google collabs to perform the scenarios so the participant can execute the task online. 
The scenarios consist of:
\begin{enumerate}
    \item The first scenarios will ask the participants to use the BOPI tool to find the article on a specific topic they would be interested in "CSV" format from \href{https://depositonce.tu-berlin.de/}{Depositonce} and listing the dataset inside the repository before the participants download the dataset and store the dataset in the folder, they should preview it at the head of the file and end of the file. Lastly, transform it to JSON and profile the data to check if the dataset is correct.
    \item The second scenario will have the same task as the first scenario to Exploring research data without the support of BOPI.
\end{enumerate}

\section{Results and Discussion}
\label{results}
In this section, we present the results and discuss the evaluation of our proposed approach in tree different dimensions: Satisfaction, Effectiveness and Efficiency.

\subsection{Effectiveness}
\begin{figure}[h!]
  \includegraphics[scale=0.5]{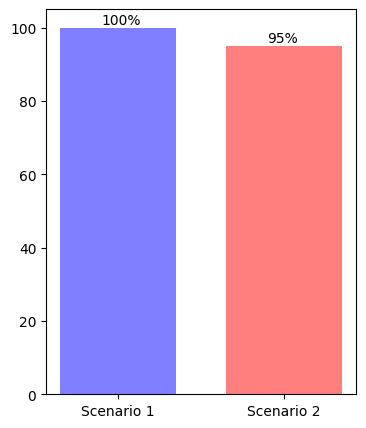}
  \caption{Completion rates of using BOPI (Scenario 1) and without BOPI (Scenario 2)}
  \label{fig:Ebild}
\end{figure}

The Effectiveness is measured by calculating the completion rate give by equation \ref{fig:completionRate} in Picture \ref{fig:Ebild}.

\begin{table}[H]
\begin{tabular}{|c|c|c|}
\hline
\textbf{Participant}                                                & \textbf{Task 1} & \textbf{Task 2} \\ \hline
\textbf{1}                                                          & 1               & 1               \\ \hline
\textbf{2}                                                          & 1               & 0               \\ \hline
\textbf{3}                                                          & 1               & 1               \\ \hline
\textbf{4}                                                          & 1               & 1               \\ \hline
\textbf{5}                                                          & 1               & 1               \\ \hline
\textbf{6}                                                          & 1               & 1               \\ \hline
\textbf{7}                                                          & 1               & 1               \\ \hline
\textbf{8}                                                          & 1               & 1               \\ \hline
\textbf{9}                                                          & 1               & 1               \\ \hline
\textbf{10}                                                         & 1               & 1               \\ \hline
\textbf{11}                                                         & 1               & 1               \\ \hline
\textbf{12}                                                         & 1               & 1               \\ \hline
\textbf{13}                                                         & 1               & 1               \\ \hline
\textbf{14}                                                         & 1               & 1               \\ \hline
\textbf{15}                                                         & 1               & 1               \\ \hline
\textbf{16}                                                         & 1               & 1               \\ \hline
\textbf{17}                                                         & 1               & 1               \\ \hline
\textbf{18}                                                         & 1               & 1               \\ \hline
\textbf{19}                                                         & 1               & 1               \\ \hline
\textbf{20}                                                         & 1               & 1               \\ \hline
\textbf{Success}                                                    & 20              & 19              \\ \hline
\textbf{\begin{tabular}[c]{@{}c@{}}Completion\\ Rates\end{tabular}} & 100\%           & 95\%            \\ \hline
\end{tabular}
\vspace*{5mm}
\caption{Completion rates tables}
\end{table}

\subsection{Efficiency}

\begin{table}[H]
\begin{tabular}{|c|c|c|}
\hline
\multicolumn{1}{|l|}{}                                                                & \textbf{Task 1} & \textbf{Task 2}              \\ \hline
\textbf{P1}                                                                           & 327             & 288                          \\ \hline
\textbf{P2}                                                                           & 751             & 797                          \\ \hline
\textbf{P3}                                                                           & 296             & 240                          \\ \hline
\textbf{P4}                                                                           & 211             & 437                          \\ \hline
\textbf{P5}                                                                           & 228             & 310                          \\ \hline
\textbf{P6}                                                                           & 310             & 300                          \\ \hline
\textbf{P7}                                                                           & 235             & 257                          \\ \hline
\textbf{P8}                                                                           & 481             & 354                          \\ \hline
\textbf{P9}                                                                           & 197             & 233                          \\ \hline
\textbf{P10}                                                                          & 380             & 430                          \\ \hline
\textbf{P11}                                                                          & 161             & 317                          \\ \hline
\textbf{P12}                                                                          & 182             & 224                          \\ \hline
\textbf{P13}                                                                          & 229             & 374                          \\ \hline
\textbf{P14}                                                                          & 281             & 560                          \\ \hline
\textbf{P15}                                                                          & 149             & 250                          \\ \hline
\textbf{P16}                                                                     & 170             & 253                          \\ \hline
\textbf{P17}                                                                          & 218             & 342                          \\ \hline
\textbf{P18}                                                                          & 209             & 224                          \\ \hline
\textbf{P19}                                                                          & 398             & 405                          \\ \hline
\textbf{P20}                                                                          & 246             & 249                          \\ \hline
\textbf{Avg. TOT*}                                                                    & 283             & 342                          \\ \hline
\textbf{\begin{tabular}[c]{@{}c@{}}Timed-Based Efficiency\\ (goals/sec)\end{tabular}} & 0.0041          & 0.0031                       \\ \hline
\textbf{Overall Relative Efficiency}                                                  & 100\%           & \multicolumn{1}{l|}{88.35\%} \\ \hline
\end{tabular}
\vspace*{5mm}
\caption{Time on task in seconds}
\label{tbl:table3}
\end{table}

\subsection{Satisfaction}

We have used System Usability Scale (SUS)~\cite{metrics} to measure the satisfaction. It is a mostly employed strategy for measure satisfaction and is considered an ideal scale for the participants due to simplicity and quick understanding by the participants.   

The calculation of System Usability Scale (SUS) Score for each participants will need several steps as following points:
\begin{enumerate}
    \item Convert the User ratings into the number for each of the 10 questions:
    \begin{itemize}
        \item Strongly Disagree: 1 point
        \item Disagree: 2 points
        \item Neutral: 3 points
        \item Agree: 4 points
        \item Strongly Agree: 5 points
    \end{itemize}
    \item Calculate the SUS Score
    \begin{itemize}
        \item X = Odd-numbered questions - 5 
        \item Y = 25 - Even-numbered-questions
        \item SUS SCore = (X+Y) * 2.5
    \end{itemize}
   \item Calculate the average of all user's Scores:
   \begin{itemize}
       \item Avg. Score = User 1 score + User 2 score + [etc.] / number of users
   \end{itemize}

\end{enumerate}

\begin{figure}[h!]
  \includegraphics[scale=0.5]{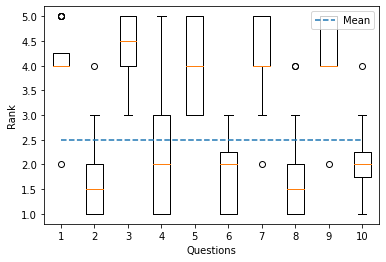}
  \caption{Descriptions of questionnaire}
  \label{fig:qBild}
\end{figure}

\begin{figure}[h!]
  \includegraphics[scale=0.5]{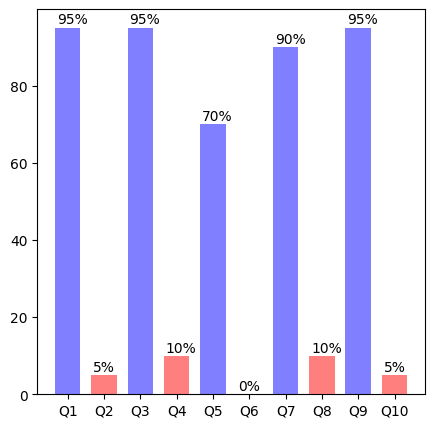}
  \caption{Percentage of Agree based on questions}
  \label{fig:ABild}
\end{figure}

\begin{table}[h!]
\begin{tabular}{l|c|c|l}
\cline{2-3}
                                                          & \textbf{Task 1} & \textbf{Task 2} &  \\ \cline{1-3}
\multicolumn{1}{|l|}{\textbf{Easier to explore the data}} & 4.2 (90\%)      & 3.9 (70\%)      &  \\ \cline{1-3}
\multicolumn{1}{|l|}{\textbf{Accomplish task quicker}}    & 4.3 (95\%)      & 3.8 (70\%)      &  \\ \cline{1-3}
\multicolumn{1}{|l|}{\textbf{Ease of use}}                & 4.5 (95\%)      & 4.4 (95\%)      &  \\ \cline{1-3}
\multicolumn{1}{|l|}{\textbf{Overall}}                    & 4.3             & 4.0             &  \\ \cline{1-3}
\end{tabular}
\vspace*{5mm}
\caption{Mean Task Ratings and Percent Agree}
\end{table}

\begin{table}[]

\centering
\begin{tabular}{|l|l|}
\hline
\textbf{Participant}              & \textbf{SUS Score}              \\ \hline
1                                 & 82.5                            \\ \hline
2                                 & 47.5                            \\ \hline
3                                 & 70                              \\ \hline
4                                 & 77.5                            \\ \hline
5                                 & 75                              \\ \hline
6                                 & 67.5                            \\ \hline
7                                 & 90                              \\ \hline
8                                 & 90                              \\ \hline
9                                 & 100                             \\ \hline
10                                & 80                              \\ \hline
11                                & 82.5                            \\ \hline
12                                & 72.5                            \\ \hline
13                                & 72.5                            \\ \hline
14                                & 97.5                            \\ \hline
15                                & 90                              \\ \hline
16                                & 82.5                            \\ \hline
17                                & 75                              \\ \hline
18                                & 70                              \\ \hline
19                                & 82.5                            \\ \hline
20                                & 87.5                            \\ \hline
\multicolumn{2}{|l|}{AVG = SUS Scores /  total participants = 79.6} \\ \hline
\end{tabular}
\vspace*{5mm}
\caption{SUS Scores for all participants for first scenario}
\label{tbl:table1}
\end{table}

\subsection{Summary}

\begin{table}[H]
\begin{tabular}{|c|c|c|c|c|}
\hline
\textbf{Task} & Completion Rate & Error Rates & Time on Task & SUS \\ \hline
1             & 100\%           &   40\%    & 283          &      4.3        \\ \hline
2             & 95\%            &    35\%   & 330          &      4.0        \\ \hline
\end{tabular}
\caption{Summary of data}
\label{tbl:tableSum}
\end{table}

\subsection{Recomendations}
This section recommendation provides recommended changes and justifications driven by the participant success rate, behaviors, and comments. Each recommendation includes a severity rating. The following recommendations will improve the overall ease of use and address the areas where participants experienced problems or found the interface/information architecture unclear. 

\begin{enumerate}
    \item  \textbf{Change:}  Add easier and clearer syntax to avoid confusion.
    
           \textbf{Justification:} Participants for task 1 produced 40\% error rates where the errors come from the given wrong syntax. They recommended including orderly syntax to increase ease of use percentage and avoid giving the wrong input in parameters.
           
           \textbf{Severity:} Low
    \item  \textbf{Change:}  Fix the the ID parameter.
    
           \textbf{Justification:} During the test, most participants input the wrong form of the ID from the table into the parameter. Mostly, they input the ID without double-quotes into the parameter where it produces the error. Some of them were confused because they thought that they had given the correct input. In the end, they suggested fixing the parameter from String to number without double-quotes.
           
           \textbf{Severity:} Medium
\end{enumerate}

\section{Use cases}
\label{useCase}
Berlin University Alliance has widely employed DSpace Systems as Research Data Repository. For instance, the SZF(Servicezentrum Forschungsdatamanagement) at the Technical University has provided “Depositonce” (depositonce.tu-berlin.de) as a repository for research data and publication at TU Berlin, which is based on Dspace Systems. This repository has approximately more than 9000 publications with its documents, such as articles, metadata, theses, or data sets. The other example of repository for research data and publication is “Refubium” (https://www.fu-berlin.de/sites/refubium) as the Freie Universität Berlin institutional repository. 

\subsection{Looking for Experts on Climate Change}
It is possible to use BOPI to explore the metadata in different ways. For example, once BOPI is configured to search for data at Berlin University Alliance, we can use the metadata about the research data to compile a list of top authors on the topic desired.

\begin{lstlisting}[style=mypython,caption={Exploring metadata}]
import bopi
bopi.search("Climate Change")["author"].value_counts()
\end{lstlisting}

\begin{figure}[th!]
  \includegraphics[scale=0.5]{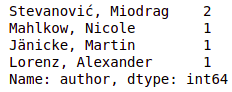}
  \caption{Top authors for the query 'Climate Change' }
  \label{fig:data_value}
\end{figure}

\subsection{Reproducible Experiments}

\begin{lstlisting}[style=mypython,caption={Example of how to automatically make download of data into an experiment}]
from sklearn.cluster import KMeans
import bopi

## Using BOPI to integrate data from a DSpace 
## repository about climate change.
DATA=bopi.repository("11303/10989.2")
    .datasets().get('name_of_file.csv').to_json()

# Clustering time and temperature using KMeans
kmeans = KMeans(n_clusters=2)
    .fit(DATA[["time","temperature"]])
print(kmeans.labels_)
print(kmeans.cluster_centers_)
\end{lstlisting}

\section{Discussion}
\label{discussion}
In general, participant's task performance showed that exploring data using the BOPI interface was easy and without significant obstacles. Most participants completed these two tasks within a reasonable time duration (averaging about 5 minutes) and did not require extensive support from the technical supporter. The results indicate that there is a significant difference in the aspect of usability that we can see from the effectiveness result based on metrics completion rate and error rate. It has proven that most of the participants could manage to finish the first scenario, and only 1 participant could not manage to finish the second scenario because there was a point where the participant had to search in google to finish the task, and this 1 participant failed to find the way to solve the task and decided to give up. 
For the efficiency aspect, the data suggests the average time spent per task where it has proven the smaller amount of time taken to complete the first scenario rather than the second scenario where the user spent most of the time to find the data. The satisfaction aspect measured by SUS Scores shown in table 5, where its score surpasses the average value of standard SUS Scores(68).  

These results build on existing evidence of Volentine et al. ~\cite{improve}. The usability test method that DATAOne has performed by identifying usability problems, collecting quantitative data to be analyzed, and determining user satisfaction. We have covered some uncovered usability issues found by the DATAOne usability test to achieve a better result. Compared with the previous study, we have performed the pre-evaluation to make sure that we will find uncovered issues before conducting post-evaluation. Additionally, this pre-evaluation has a purpose of avoiding the mistake during evaluation such as the error function or unreadable scenarios. Another method that we have used to evaluate the participants' satisfaction by using SUS scores seems useful since it includes only ten statements and gives us the average scores to measure usability and learnability.

In contrast, the results' generalizability is limited by the producing output that takes more time than we have estimated. Most of the participants encountered long responses as they wanted to see the result. We are trying to cover this issue by fixing the code to make the response faster. Other important participant confusion issues were the ID parameter that the user prefers ID as the number than string characters and unclear syntax. Moreover, it is beyond this study's scope to involve more than 20 participants because of the budget issues and hard to find the participants.

Despite the usability issues revealed from the evaluation, participants valued the tool's utility and were interested in following the development of our tool. Some new suggestions from participants have been noted, such as: adding new features to read more types of data and cleaning the data. Most participants agreed that the BOPI interface tools could improve the usability of data reuse among the researchers.

\section{Conclusion and future work}
\label{conclusion}
In sum, our primary goal addresses the following problems: a) Monolingualism, i.e., Research data at data repositories being shared in only one language, which would make data less findable; and b) Embeddability,  i.e., Create a mechanism to embedding research data into a scientific experiment, improving the reproducibility of experiments.

This paper's objective was to improve the usability in research data repositories of DSpace and prove that our programming interface will improve the usability aspects for researchers in reproducing experiments. Participants with experienced programming users were recruited for the evaluation, and they performed well on exploring data with our tool and without our tool. This evaluation collected completion rates, error rates, mean task rating, time on task, satisfaction scores, participant comments, and observational notes to identify workflow and usability issues. 
The results revealed that unclear syntax in tools and the delay of output caused some confusion among participants; specific improvements for features (e.g., adding data cleaning feature, reading various data format features, and change to easier syntax) could be a good improvement for this tool.
Furthermore, the usability evaluation also showed that the translation metadata, preview of the data to be downloaded, and automatic data transformation make data more reusable and accessible. 
The evaluation's limitation was the limited number of participants, which could not lead to statistically significant results. The other issue was the lack of datasets provided by the DSpace repository for depositonce. Nevertheless, the evaluation is part of an iterative design process, and its purpose to identify usability issues and provide suggestions through the scenarios and feedback. 
As we tested the tool, we also noticed the new usability issues with new participants, which is expected in usability research practice.

We have identified possible few limitations in our solution, which due to the scope of this project, were not addressed, such as:
\begin{itemize}
    \item Translation: we have opted to use Google translation API, which is addressed to general texts and can give the wrong interpretation of scientific words' meaning. Moreover, translation for languages e.g., English, has more accurate results than resourceless languages e.g., Indonesian or Chinese.
    
    \item Audience: Although this tool's use is simple, we are aware that some researchers that do not have prior knowledge of Python.
    
    \item Scope: we have designed our tool to work only with DSpace systems, which is only one type of system available for this end.
    
\end{itemize}

As future work, we highlight the importance of making a Usability Assessment focused on FAIR criteria. The FAIR principles define how research output must be easy to find by humans and computers and containing sufficiently rich metadata. A unique and persistent identifier (\textbf{F}indable), metadata, and data are accessible to humans or machine reading, including authentication and authorization. (\textbf{A}ccessible), (Meta)data should use formal and shared functional language for knowledge representation and references to other (meta)data. (\textbf{I}nteroperable), data should clearly define usage licenses and come up with accurate information on the source (\textbf{R}eusable)~\cite{fair}. Our goal is to perform a structured and comprehensive mapping of the FAIR requirements on the DSpace solutions where every facet of the FAIR principle needs to be implemented to analyze the system's design of DSpace. As a methodological approach, we want to identify the representative group of users by interviewing participants. First, we want to perform semi-structured interviews with different expert and stakeholder groups and focus group sessions on gathering more insight and a more detailed understanding of user's perceptions and attitudes towards the usage of research data repositories. In a second step, we are going to assess the usability criteria of (1) user satisfaction, (2) effectiveness, and (3) efficiency in DSpace System repositories with a mixed-method approach explicitly focusing on the structural and features of DSpace.

\bibliographystyle{ACM-Reference-Format}
\bibliography{sample-base}

\appendix

\end{document}